\documentclass[english]{article}
\usepackage[T1]{fontenc}
\usepackage[utf8]{inputenc}
\usepackage[a4paper]{geometry}
\usepackage{babel}
\usepackage{soul}
\usepackage{textcomp}
\usepackage[unicode=true]
 {hyperref}

\usepackage{amsmath}
\usepackage{amssymb}
\usepackage{graphicx}
\usepackage{multirow}
\usepackage{arydshln}
\usepackage{siunitx} 

\usepackage{url}

\usepackage{subcaption}
\usepackage{makecell}
\usepackage{bm}
\usepackage{mathtools}

\usepackage{todonotes}

\newcommand{\doa}{\bm \theta} 

\newcommand{\target}{\bm \theta_\mathrm{t}}
\newcommand{\targetNormalized}{\bar {\bm \theta}_\mathrm{t}}

\newcommand{\gt}{\bm s}
\newcommand{\estimated}{\hat{\bm s}}

\newcommand{\bformer}{\hat{\bm s}_{\mathrm{BF}}}
\newcommand{\ambimix}{\bm X}

\newcommand{\added}[1]{{\color{black} #1}}
\newcommand{\deleted}[1]{}

\usepackage{lineno}

\makeatletter
\newcommand{\lyxaddress}[1]{
	\par {\raggedright #1
	\vspace{1.4em}
	\noindent\par}
}

\@ifundefined{date}{}{\date{}}
\makeatother

\begin{document}
\title{Direction Specific Ambisonics Source Separation with End-To-End Deep Learning}
\author{Francesc Llu\'is$^{1)}$, Nils Meyer-Kahlen$^{2)}$, Vasileios Chatziioannou$^{1)}$, Alex Hofmann$^{1)}$}
\maketitle

\lyxaddress{$^{1)}$ Dept. of Music Acoustics, University of Music and Performing Arts, 1030 Vienna, Austria}

\lyxaddress{$^{2)}$ Dept. for Signal Processing and Acoustics, Aalto University, Otakaari 5, 0215 Espoo, Finland}

\begin{abstract}
Ambisonics is a scene-based spatial audio format that has several useful features compared to object-based formats, such as \added{efficient} whole scene rotation and versatility. However, it does not provide direct access to the individual source signals, \deleted{and}\added{so that} these have to be separated from the mixture when required. Typically, this is done with linear spherical harmonics (SH) beamforming. In this paper, we explore deep-learning-based source separation on \added{static} Ambisonics mixtures. In contrast to most source separation approaches, which separate a fixed number of sources of specific sound types, we focus on separating arbitrary sound from specific directions. Specifically, we propose three operating modes that combine a source separation neural network with SH beamforming: refinement, implicit, and mixed mode. We show that a neural network can implicitly associate conditioning directions with the spatial information contained in the Ambisonics scene to extract specific sources. We evaluate the performance of the three proposed approaches and compare them to SH beamforming on musical mixtures generated with the musdb18 dataset, as well as with mixtures generated with the FUSS dataset for universal source separation, under both anechoic and \deleted{reverberant}{room} conditions. Results show that the proposed approaches offer improved separation performance and spatial selectivity compared to conventional SH beamforming.

\end{abstract}
\section{Introduction}

\begin{figure}[t]
    \center
    \label{}
    \includegraphics[]{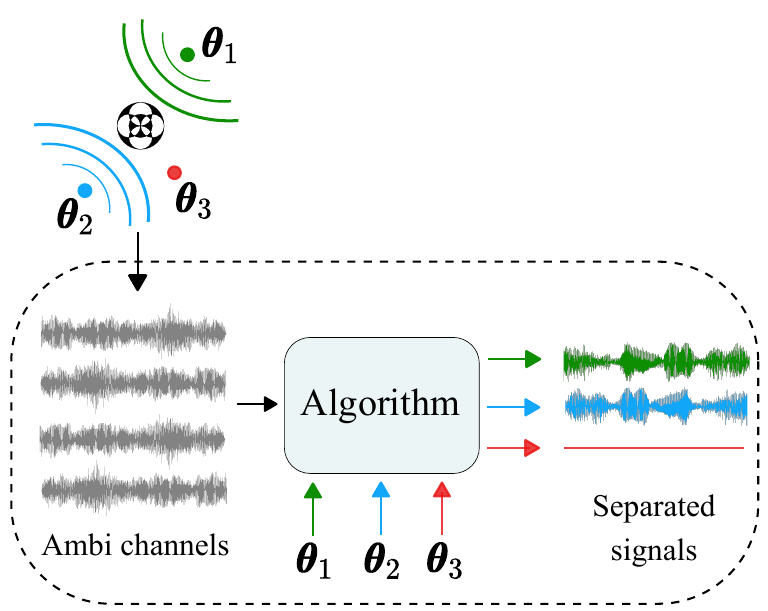}
    \caption{Illustration of an algorithm separating the sound sources from a raw Ambisonics mixture given the directions of interest. In this case, two sources (blue and green dots) located at directions $\bm  \theta_1$ and $\bm \theta_2$ are emitting sound, while no sound is coming from direction $\bm \theta_3$.}
    \label{fig:teaser}
\end{figure}

Ambisonics is a scene-based spatial audio format \cite{zotter_ambisonics_2019} that is widely adopted in immersive and virtual reality audio applications. It is based on representing the audio scene in the spherical harmonics (SH) domain, where each Ambisonics channel represents one SH component out of an order-limited set. This offers a universal framework for recording, transmitting and reproducing spatial audio material, with several useful features such as efficient whole scene rotation that does not depend on the number of sources in the scene, and independence between capturing and reproduction setup. This versatility comes at the cost of not having access to isolated signals of individual sound sources in the scene, as is the case in object-based spatial audio formats. Yet, access to individual sound sources or the ability to extract sound from specific directions is often required, for example, to apply modifications to the scene, or to perform analysis or further processing on individual sources. \added{Apart from post-processing of Ambisonics music or natural scene recordings, source separation algorithms can be used for enhancing certain sound sources like human speakers from a complete scene captured with head-worn arrays \cite{guiraud_2023}. The presented algorithm would be applied after converting such a capture to the Ambisonics domain \cite{ahrens_spherical_2021, mccormack2022parametric}.}

The most conventional approach for extracting individual sources from a\added{n} \added{Ambisonics} scene is to exploit the spatial information by applying SH beamforming. To do so, sound from a target direction is extracted by linear combination of the Ambisonics channels. Through the years, different beamformer designs have been presented \cite{teutsch2007modal, rafaely_fundamentals_2014, jarrett_theory_2017}, including both signal-independent and signal-dependent variants.

A completely different family of methods for extracting sound sources from a mixture is based on source separation techniques, which operate in the time-frequency domain. Source separation has been extensively studied for single-channel mixtures, but approaches have been proposed to separate sources from multichannel recordings as well, for example \cite{nugraha2016multichannel, ozerov2009multichannel}. 
Although using specifically Ambisonics mixtures for multichannel source separation is still relatively rare, some methods have been presented. Epain et al. \cite{epain2012independent} applied independent component analysis and Hafsati et al. \cite{hafsati_sound_2019}  used local Gaussian modelling to perform source separation on Ambisonics signals. Also, several authors have proposed the use of multichannel non-negative matrix factorization \cite{nikunen2018multichannel}, and non-negative tensor factorization in several variants \cite{munoz2021ambisonics, mitsufuji2021evan, guzik2022wishart}, operating on Ambisonics signals. In comparison to beamforming where a target direction is selected, all these source separation techniques extract sound sources of a certain type from the mixture, the number of sources needs to be known, and the complete scene is decomposed into all components. In addition, the separation stage is computationally expensive and time demanding.

Concurrently, advances in deep learning have led to large improvements in audio source separation in comparison to other methods \cite{mitsufuji2021music, cobos_overview_2022}.  
As a logical development, some approaches that combine deep learning methods and spatial processing through beamforming have already been proposed. For example, in \cite{bosca2021dilated, ochiai_beam-tasnet_2020} a neural network is used to inform a beamformer that predicts frequency-dependent signal and noise covariance matrices to perform spatial filtering. Although encouraging results are obtained with such learning-based approaches, they are trained and tested assuming a deterministic number of sources in the mixture and closed domain sound types, such as speech signals. In contrast, our approach aims to separate any desired number of sources and type of signals. A related approach uses multichannel audio recordings and a neural network to separate speech signals in the horizontal plane \cite{jenrungrot2020cone}, but it operates on microphone array data instead of Ambisonics signals and localizes sound sources itself, rather than separating sound from a specified direction.

In this paper, we adopt a data-driven approach to the task of extracting signals from specific directions given an Ambisonics mixture (as illustrated in \deleted{Fig.}\added{Figure}~\ref{fig:teaser}). Specifically, we explore three operating modes that involve end-to-end deep learning, i.e. from waveform to waveform, and SH beamforming: 1) refinement mode, 2) implicit mode, and 3) mixed mode. In refinement mode, the deep neural network is used solely for the refinement of the single channel SH beamformer output, pointing to a target direction. In implicit mode, the Ambisonics mixture is directly provided to the network and the target direction is used to condition the network output. In mixed mode, both the Ambisonics mixture and the beamformer output are provided to the network, while the target direction is also used to condition the network output. The aim of the study is then to analyze the performance of the three modes and to compare them to conventional SH beamforming, for different Ambisonics orders. To do so, we assess the source separation performance in the direction of the sources and the capability of the methods to predict silence in regions where no source is placed, i.e. the spatial selectivity. These evaluations are performed for anechoic and room conditions, and considering musical mixtures and universal mixtures, composed by unknown number of sources from an open domain of sound types. The performance is reported considering Ambisonics mixtures with order between one and four. The code\footnote{\url{https://github.com/francesclluis/direction-ambisonics-source-separation}} and listening examples \cite{listening_examples} accompanying the paper are available online.

The paper is organized as follows. Section \ref{sec:background} provides background on Ambisonics and SH beamforming, which serves as a baseline in our experiments. Section \ref{sec:approach} details the neural network architecture and the training procedure. The datasets used in the experiments are described in Section \ref{sec:datasets}. Section \ref{sec:evaluation} presents the evaluation metrics and the obtained results. Section \ref{sec:discussion} discusses the results and Section \ref{sec:conclusion} concludes the paper.

\section{Background}
\label{sec:background}

\subsection{Ambisonics}

In an ideal, instantaneous Ambisonics mixture, $K$ far-field sound sources $s_k(t)$ for $k \in [1,..., K]$, placed at directions $\bm \theta_k$ are represented as
\begin{align}
    \bm \chi_N(t) = \sum_{k=1}^K s_k(t) \bm y_N\bm (\bm \theta_k),
    \label{eq:encoding}
\end{align}
where $\bm y_N(\bm \theta) = \begin{bmatrix}
Y_0^0(\bm \theta) & Y_1^{-1}(\bm \theta)  & Y_1^{0}(\bm \theta) \ \hdots & Y_N^{N}(\bm \theta) \end{bmatrix}^\top$ is a vector of real-valued spherical harmonics, defined as
\begin {align}
Y_n^m(\doa) = &N_{nm} P_n^m(\sin \vartheta)	\begin{cases}
\sqrt{2}\sin (|m|\phi)& m < 0 \\
1 & m=0\\
\sqrt{2}\cos (m \phi)  & m > 0
\end{cases},\\
&N_{nm} =  \sqrt{\frac{(2n+1)(n-m)!}{4 \pi (n+m)!}},
\end{align}
evaluated at the direction $\doa = [\phi, \vartheta]$, where \added{0 $\leq$} $\phi$ \added{$<2 \pi$} is the azimuth and \added{$0 \leq$} $\vartheta$ \added{$\leq \pi$} is the zenith angle. $N$ is the maximal Ambisonics order and $P_n^m$ are the associated Legendre polynomials. In acoustics, it is common to call the index $0 \leq n \leq N $ the \emph{order} and $-n \leq m \leq n$ the \emph{degree} of each SH component. In the full set of $(N+1)^2$ signals, each channel of $\bm \chi_N$ represents one spherical harmonic.

A convolutive mixture can be represented using SH domain directional room impulse reponses (DRIR), sometimes also called Ambisonics room impulse responses (ARIR), $\Breve{\bm h}_k(t) \in \mathbb{R}^{(N+1)^2}$, which describe the transfer\deleted{-}path between each source and an ideal SH receiver
\begin{align}
    \bm \chi_N(t) = \sum_{k=1}^K s_k(t) * \Breve{\bm h}_k(t).
    \label{eq:convolutive}
\end{align}

\subsection{SH Beamforming}

To spatially separate sound from an Ambisonics mixture, SH beamforming uses a linear combination of the Ambisonics channels. Ideally, a beam pattern that is constant over frequency can be achieved with a real, frequency-independent weight vector $ \bm d \in \mathbb{R}^{(N+1)^2}$ such that

\begin{align}
    \hat{s}_{\mathrm{BF}}(t) = \bm d^\top \bm \chi_N(t),
    \label{eq:bf}
\end{align}

\noindent where $\hat{s}_{\mathrm{BF}}$ is the output of the beamformer, $[.]^\top$ is the transpose operator, and $\bm \chi_N$ is the Ambisonics mixture of order $N$. Note that the weights $\bm d$ can be chosen arbitrarily and their choice determines the direction and shape of the implied \deleted{beampattern}{beam-pattern}.  

In this work, we use two common frequency- and signal-independent SH beamformers as baselines: \added{the beamformer with maximal directivity index (max-DI) \cite{rafaely_fundamentals_2014, jarrett_theory_2017}, which is sometimes also referred to as plane-wave decomposition, and the beamformer with maximal energy vector (max-$\bm r_\mathrm{E}$)} \cite{zotter_ambisonics_2019}. \added{As signal-independent beamformers,} they extract sound from a specified target direction $\target$, and the shape of the beam is derived from the global optimizatin criteria, DI, and $\bm r_\mathrm{E}$ vector length. \deleted{The max-DI beam represents the beam pattern with maximal directivity index while the max-$\bm r_\mathrm{E}$ beam represents the beam pattern with maximal $\bm r_\mathrm{E}$-vector length.} The two objectives are given by 

\begin{align}
    \text{DI} = 10 \log \frac{4\pi g^2(\target)}{\int_{\bm \theta} g^2(\bm \theta) d\bm \theta}, \qquad
      \bm r_{\mathrm{E}} =  \frac{\int_{\bm \theta} g^2(\bm \theta) \bm \theta \added{d \bm \theta}}{\int_{\bm \theta} g^2(\bm \theta)\added{d\bm \theta}} ,
\end{align}

\noindent respectively, where $g(\bm \theta)$ is the pattern of the beamformer evaluated at the direction $\bm \theta$. It is obtained by evaluating
\begin{align}
    g(\bm \theta) = \bm d^\top \bm y_N(\bm \theta).
\end{align}

\begin{figure*}
    \centering
    \includegraphics[width = 17cm]{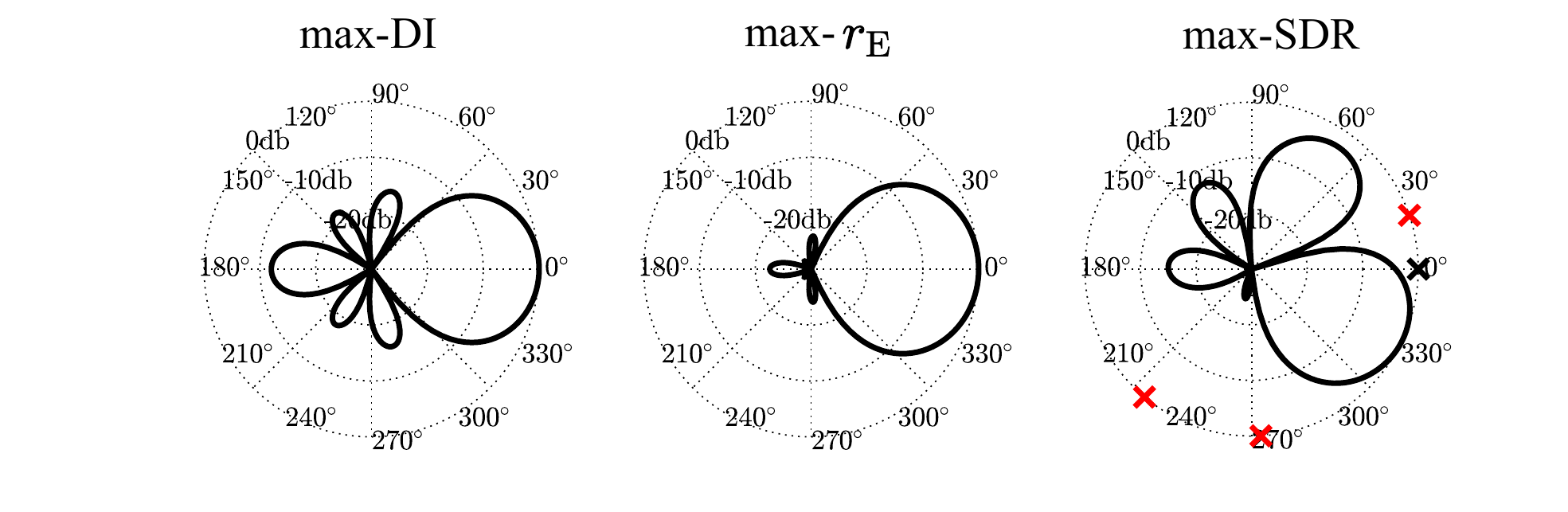}
    \caption{Signal-independent max-DI and max-$\bm r_\mathrm{E}$, and signal-dependent max-SDR Beamformer \added{pointing to a source at ($0^\circ$, $0^\circ$)} for a maximal order of $N = 3$. The black cross symbolizes the source and the red crosses symbolize interferers. 30 dB dynamics are shown.}
    \label{fig:bf}
\end{figure*}

Weights that optimize these global criteria are given by
\begin{align}
    \bm d_{\mathrm{max\text{-}DI}} &= \bm y_N(\target), \quad \bm d_{\mathrm{max\text{-}\bm r_{\mathrm{E}}}} = \text{diag}_N(w_n) \bm y_N(\target),  
\end{align}
\noindent \added{with the max-$\bm r_\mathrm{E}$ order weights, which can be approximated by \cite[p. 188]{zotter_ambisonics_2019}}
\begin{align}
    w_n \approx P_n\Big(\frac{{\cos (137.9^\circ)}}{N+1.51}\Big),
\end{align}
in which $P_n$ is the Legendre function of n-th order \cite{zotter_ambisonics_2019}, \added{and $\text{diag}_N(w_n)$ denotes expanding the order weights $w_n$ to a diagonal matrix, with one weight for all the elements corresponding to one order $n$.} See \deleted{Fig.}\added{Figure}~\ref{fig:bf} for the corresponding beam patterns. While the max-DI pattern has a narrow main lobe at the cost of significant side lobes, the \added{max-}$\bm r_{\mathrm{E}}$ pattern offers a compromise between \deleted{the two}\added{main lobe width and side lobe strength}.

In addition, we use another SH beamformer for comparison, which we refer to as the max-SDR beamformer. Given a particular ground truth signal of length $T$, now written as a vector $\bm s = \begin{bmatrix} s(0), s(1), ..., s(T-1) \end{bmatrix}^\top$, the max-SDR beam represents the beam pattern that extracts the signal $\gt$ with the \deleted{maximally}\added{maximum} possible \deleted{signal}\added{source}-to-distortion (SDR)\added{\cite{vincent2007first}} ratio  through SH beamforming. The SDR between a reference signal $\gt$ and an estimated signal $\hat{\bm s}$ is defined as:
\begin{equation}
\textrm{SDR}(\bm s, \hat{\bm s}) = 10\log_{10}\left(\frac{\|\bm  s\|^2}{\| \bm s - \hat{\bm  s}\|^2}\right).
\end{equation}

The maximal SDR is achieved by finding the minimum squared error between ground truth and estimated sources. This is equivalent to finding the \added{minimum mean squared error} (MMSE) beamformer \cite[p.446]{van_trees_detection_2002}, given full knowledge of the source signal. The coefficients are found as

\begin{align}
    \bm d_{\text{max-SDR}} = \bm C^{-1} \bm X_N^\top \bm s, \label{eq:maxSDR} 
\end{align}
where $\bm C= (\bm X_N^\top \bm X_N)$ is the spatial covariance matrix of the input signal of length  $T$, which is stacked into a matrix  $\bm X_N = \begin{bmatrix} \bm  \chi_N(0), \bm \chi_N(1), ..., \bm \chi_N(T-1) \end{bmatrix}^\top$.

As opposed to the other beamformers, the max-SDR pattern is signal-dependent. The approach is not directly applicable in practical situations, as the \deleted{the} ground truth signal $\gt$ is obviously not available. Also, \deleted{the white noise gain might be high}\added{some channels might be amplified excessively, which} would introduce noise. In our investigation, the use of the max-SDR beamformer is of theoretical interest. It serves as an upper bound, to see which separation is \deleted{maximally}\added{maximum} possible by frequency-independent spatial processing alone, assuming that the ground truth source signal is known \deleted{as oracle information}. As seen in \deleted{Fig.}\added{Figure}~\ref{fig:bf}, the max-SDR beamformer will tend to place zeros in the directions of interfering sources, if possible.

\section{Proposed Approach}
\label{sec:approach}

\begin{figure*}[t]
    \center
    \label{}
    \includegraphics[]{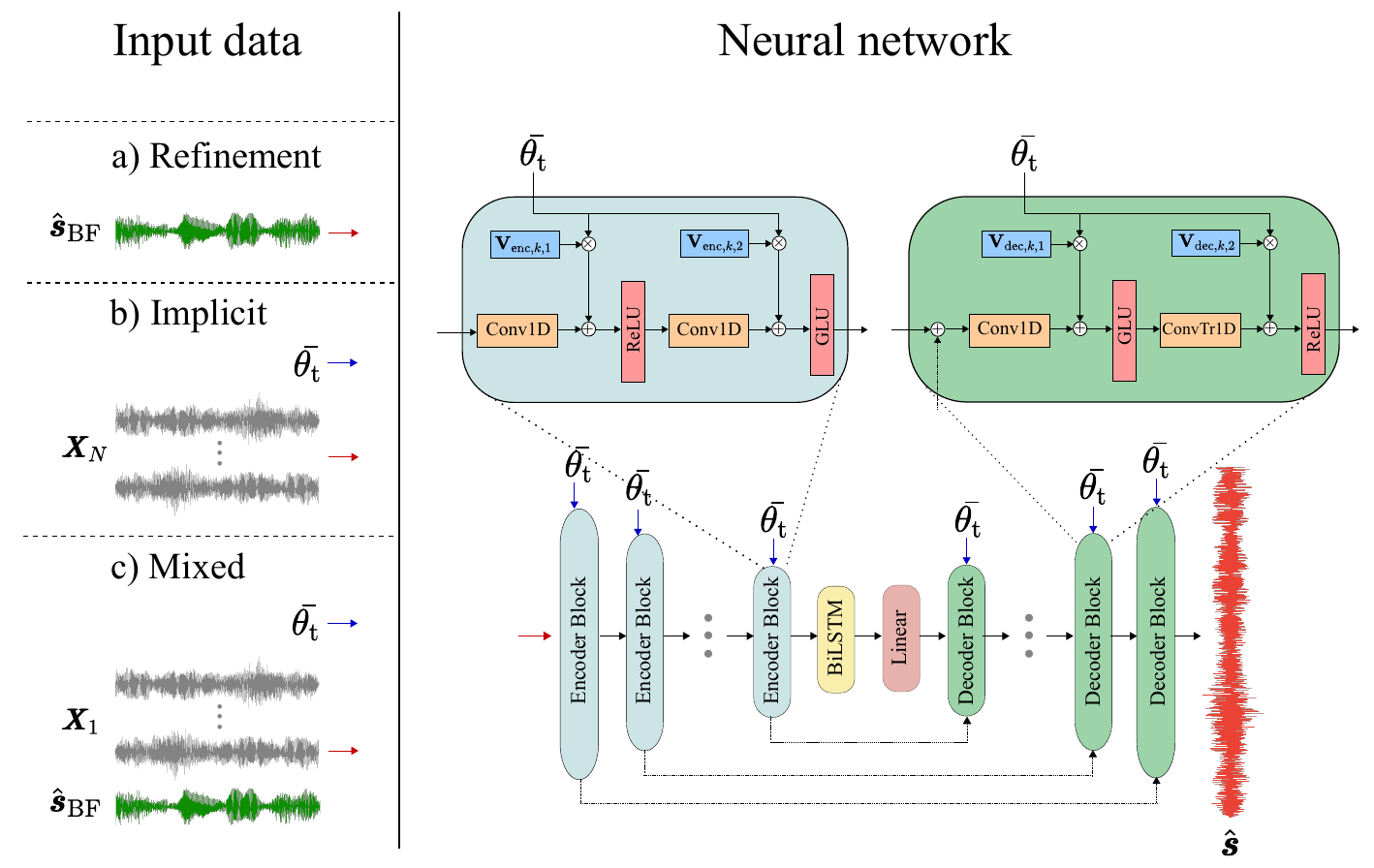}
    \caption{Left: Input data for the following operating modes: a) refinement, b) implicit, c) mixed. Right: Overview diagram of the neural network. $\ambimix_N$ refers to the n-th order Ambisonics mixture, $\ambimix_1$ is the first order Ambisonics mixture, $\bar{\target}$ is the scaled target direction, $\bformer$ corresponds to the output of a max-$\bm r_\mathrm{E}$ beamformer at the target direction, and $\hat{\bm s}$ is the estimated separated signal.}
    \label{fig:main_diagram}
\end{figure*}

We propose the use of deep learning to estimate the signal $\gt$ from a specific target direction $\target$, given the raw Ambisonics mixture $\ambimix_N$ as input signal. To this end, we explore three different operating modes which involve a deep neural network and SH beamforming.

The first is refinement mode, where the goal is to find a function $f_1$ with the structure of a neural network such that $f_1(\hat{\bm s}_{\mathrm{BF}}) = \gt$. In this case, the neural network is used solely to refine the single channel beamformer output, which points to the target direction. Note that the neural network is not informed about the target direction and only relies \deleted{in}{on} the beamformer output to enhance the beamforming operation. In addition, the beamformer output is normalized before entering the neural network as we want the proposed method to be independent of the beamformer output gain.

For the second mode, implicit mode, the goal is to find a function $f_2$ such that $f_2(\ambimix_N, \target) =  \gt$. In this case, the raw Ambisonics signal and the target direction are directly passed to the neural network. The training objective then forces the neural network to implicitly perform the whole beamforming operation, by learning the correspondence between the spatial information contained in the Ambisonics mixture and the conditioning direction to then perform source separation.

The third one is mixed mode, where the goal is to find a function $f_3$ such that $f_3(\ambimix_1, \target, \hat{\bm s}_{\mathrm{BF}}) = \gt$. In this case, the output of the SH beamforming is concatenated to the first order Ambisonics mixture as an extra channel. Note that the output of the SH beamforming is computed with the corresponding order but the Ambisonics mixture given to the neural network is always fixed to first order. This decision was made based on preliminary observations, similar to \cite{lluis_deep_2022}, where an increase in order of the Ambisonics input, did not substantially improve the overall performance, whereas an increase of the beamforming order did.

For all operating modes, $\bformer$ corresponds to the output of a max-$\bm r_\mathrm{E}$ beamformer at the target direction and the function $f$ is an adapted version of the Demucs neural network architecture \cite{defossez2019music}. 
Specifically, Demucs input channels are modified according to the Ambisonics order and for implicit and mixed mode a global conditioning approach \cite{jenrungrot2020cone} is used to guide the separation according to the target direction. Demucs was originally designed to separate four well defined instrument types from single-channel mixture signals. Hence, the original Demucs architecture outputs as many channels as known sources where each output corresponds to an instrument. In the present work, the output of the Demucs network is always a single channel corresponding to the audio at the target direction regardless of the number of sources in the mixture or the type of sound sources.

\subsection{Demucs Architecture}

Demucs \cite{defossez2019music} is a convolutional neural network that operates in the waveform domain with a U-Net-like architecture \cite{ronneberger2015u}, i.e. an encoder-decoder architecture with skip connections (see \deleted{Fig.}\added{Figure}~\ref{fig:main_diagram}). The encoder-decoder structure can learn multi-resolution features of the Ambisonics mixture in the time-space domain which enables to capture the Ambisonics channel variations at different scales in both domains. The skip connections allow to propagate low level information through the network which otherwise may be lost. In this case, information related to level and phase differences between the raw Ambisonics channels can be accessed in later decoding blocks for further source separation.
Each Demucs encoding block consists of an initial convolution operation that downsamples the input feature maps\deleted{,} by applying kernels with a size of 8 and a stride of 4, \deleted{and doubles}\added{while also increasing} the number of channels \added{by a factor of 2}. Note that the first block is an exception as it has a \deleted{fix}\added{fixed} number of output channels set to 64. Then, a Rectified Linear Unit (ReLU) activation function is applied follwed by a 1x1 convolution with a Gated Linear Unit Activation \cite{dauphin2017language} (GLU).
At the bottleneck of the network, i.e. between the encoder and decoder parts, a Bidirectional Long-Short Term Memory (BiLSTM) followed by a linear operation is applied to provide long range context.
The decoder part reverses the encoder process. Each decoding block first adds the skip connections from the encoder at the same level of hierarchy. Then, a 1x1 convolution with a Gated Linear Unit Activation (GLU) is applied. Next, a transposed convolution upsamples the feature maps\deleted{,} by applying kernels with a size of 8 and a stride of 4, \deleted{and halves}\added{while also halving} the number of channels. Finally, a ReLU activation is used. Note that the last decoding block is an exception, which neither halves the number of channels nor applies the ReLU activation.

\subsection{Conditioning}

In implicit and mixed mode, we use a global conditioning approach to inform Demucs about the target direction. Similarly to \cite{jenrungrot2020cone}, the conditioning information is inserted at each block of the Demucs network after being multiplied by a learnable linear projection $\mathbf{V}_{\cdot, q, \cdot}$. Specifically, in this case we scale the target direction $\target =  [\phi_\mathrm{t}, \vartheta_\mathrm{t}]$ with azimuth angle $\phi_\mathrm{t}\in[-\pi, \pi]$ and zenith angle $\vartheta_\mathrm{t}\in[0, \pi]$, such that the scaled target direction $\targetNormalized = [\bar{\phi}_\mathrm{t}, \bar{\vartheta}_\mathrm{t}]$ is defined as $\bar{\phi}_\mathrm{t}\in[-1, 1]$ and $\bar{\vartheta}_\mathrm{t}\in[-1, 1]$. Then the Demucs encoder and decoder takes the following expression:

\begin{align}
    \phantom{\texttt{EncodeEnc}}
    &\begin{aligned}
        \mathllap{\texttt{Encoder}_{q+1}} &= \text{GLU}(
        \mathbf{W}_{\text{encoder},q,2} \ast \text{ReLU}(\mathbf{W}_{\text{encoder},q,1} \ast \\  
        &\texttt{Encoder}_{q} + \mathbf{V}_{\text{encoder},q,1} \bar{\target}) + \mathbf{V}_{\text{encoder},q,2}  \bar{\target}
        ),
    \end{aligned}\\
    &\begin{aligned}
        \mathllap{\texttt{Decoder}_{q-1}} &= \text{ReLU}(
        \mathbf{W}_{\text{decoder},q,2} \ast^\top \text{GLU}(\mathbf{W}_{\text{decoder},q,1} \ast \\ & (\texttt{Encoder}_{q} + \texttt{Decoder}_{q}) 
         + \mathbf{V}_{\text{decoder},q,1} \bar{\target})
         \\ & + \mathbf{V}_{\text{decoder},q,2} \bar{\target}
        ).
    \end{aligned}
\end{align}

\noindent where $\texttt{Encoder}_{q+1}$ and $\texttt{Decoder}_{q-1}$ are the outputs from the $q\text{-th}$ level encoder and decoder blocks respectively. $\mathbf{W}_{\cdot, q, \cdot}$ are the 1-D kernel weights at the $q\text{-th}$ block. ReLU and GLU are the corresponding activation functions. The operator $\ast$ corresponds to the 1-D convolution while $\ast^\top$ denotes a transposed convolution operation, as commonly defined in the deep learning frameworks \cite{paszke2019pytorch}.

\subsection{Supervised Training}

For all operating modes, the network is trained in a supervised manner. The parameters of the network are optimized to reduce the $\ell_1$ loss between the estimated signal and the ground truth signal at the target direction. During training, as target direction, we randomly select one of the source directions and uniformly perturb it within a $2.5^\circ$ window. This perturbation determines the spatial selectivity of the network at the inference stage. The network is trained for 200 epochs using the Adam optimizer. The learning rate is set to $1\times 10^{-4}$ and it is reduced by a factor 0.1 after 10 epochs with no              improvement in the validation set loss. The batch size is set to 16. After the training process, we select the weights with the lowest validation loss for testing purposes. Both training and testing are conducted on a single Titan RTX GPU. The training stage takes about 18 hours while the inference takes 47 milliseconds for a single data sample (value averaged from 300 different separation predictions).

\section{Datasets}
\label{sec:datasets}

We study the performance of the proposed methods using two different datasets: the Musdb18 \cite{musdb18-hq}, which contains music signals, and the Free Universal Sound Separation (FUSS) dataset \cite{wisdom2021s}, which contains a wide range of signals from open domain sound types. In addition, for each dataset we create an anechoic and a \deleted{reverberant}{room} version. 

\subsection{Data Generation}

\subsubsection{Musdb18} We use musical signals from the Musdb18 dataset to create training, validation, and testing data. The Musdb18 dataset contains a ``train'' folder with 100 songs and a ``test'' folder with 50 songs. For each song, the dataset provides the isolated signals of the \textit{drums}, \textit{bass} and \textit{vocals} sound sources at 44.1 kHz. We use signals from 90 songs in the ``train'' folder to generate training data and the remaining 10 songs are used to generate validation data. For training and validation, a single example is created by first selecting six-second long audio segments from the isolated signals at a random time. Therefore, every isolated signal is taken from a random song for each source, so they do not necessarily come from the same piece. Then, to create an Ambisonics mixture, a random direction is assigned to each of the sources, and the audio segments are encoded to up to fourth order Ambisonics and mixed using eq. \eqref{eq:encoding}. For the generated mixtures, it is assured that all pairs of sources are at least $5^\circ$ great circle distance apart from each other. The great circle distance is the angle between two points on a sphere, defined as 
\begin{align}
    \angle (\doa_i,\doa_j) = \arctan\bm x_i^\top \bm x_j,
\end{align}
where $\bm x_i = \begin{bmatrix} \cos \phi_i \sin \vartheta_i, \sin \phi_i \sin \vartheta_i, \cos \vartheta_i \end{bmatrix}^\top$ is a normalized direction vector. In this work, we only consider mixtures with static sources.

Furthermore, in $30 \%$ of all created mixtures we force one source to be silent while we verify that the remaining mixtures contain active sources. \added{The application of this preprocessing allows the data-driven approaches to learn to provide silent output when no source is present at a given direction. This is important to assure silent output when specifying directions with no active sources during inference.} For training and validation we generate 10000 and 1000 mixtures respectively. Regarding test data generation, the same encoding is applied to generate a total of 1000 mixtures using the ``test'' folder. In this case, single examples are created using six-second long audio segments coming from the same song at the same time and no sources are silenced.

\subsubsection{FUSS} The FUSS dataset was created for universal sound separation. Universal sound separation algorithms aim to separate unknown number of sources from an open domain of sound types. To this end, FUSS contains 23 hours of single-source audio data at 16 kHz drawn from 357 classes, which are used to create mixtures of one to four sources. \added{The type of audio contained in FUSS includes natural sounds such as wind and rain, sounds of objects such as engine and alarm, and human sounds such as whistling and human voice.} FUSS provides \added{splits for} train\added{ing}, validation, and test\added{ing}\deleted{splits} with a total of 20000, 1000, and 1000 examples respectively. We use the same partition and create six-second long Ambisonics mixtures. As previously done, for each example we assign a random direction to each of the sources, and the audio segments are encoded up to fourth order Ambisonics and mixed using eq. \eqref{eq:encoding}. In this case, it is assured that all pairs of sources are active and at least $5^\circ$ great circle distance apart from each other. Note that during testing, we only consider generated mixtures that contain two or more sources.

\subsection{Room Simulation}

The anechoic Ambisonics mixtures according to \eqref{eq:encoding} are a good test case, but they are far from an actual application, as usually sound sources of interest are within an enclosed room. Therefore, performance is also studied under \deleted{reverberant}{room} conditions by incorporating directional room impulse responses (DRIR) of a small room to create a convolutional mixture, as defined in \eqref{eq:convolutive}. To create the DRIRs, sound sources are placed in a simulated room of dimensions in the range $(x, y, z) = (3\si{\m} \pm 2\si{\m}, 4\si{\m} \pm 2\si{\m}, 3\si{\m} \pm 1\si{\m})$. Early reflections are simulated using the image source method with a maximal image source order of six. Wall absorption coefficients are set in octave bands, where the reflection coefficient is determined from random octave band reverberation times $\text{RT}_f = 0.3\si{\s}\pm 0.2\si{\s}$ using Eyring's formula \cite{kuttruff_room_2017}. For late reverberation, the response is faded over the isotropic diffuse noise at $t_{\mathrm {mix}} = \sqrt{V} / 500 \ \si{\s} / \si{\m}^3 $. The decay of the noise is exponentially shaped in octave bands according to the random reverberation times. Finally, although the Ambisonics mixtures contain \deleted{reverberation caused by the room}{a simulated room}, we use the anechoic signals of the sound sources as ground truth for evaluating the separation performance.

\section{Results}
\label{sec:evaluation}

\subsection{Evaluation Metrics}

We use two different measures of performance for evaluating the proposed methods. First, we measure the separation performance using the scale-invariant source\added{-}to\added{-}distortion ratio (SI-SDR) \cite{le2019sdr}. SI-SDR between a signal $\gt$ and its estimate $\hat{\bm s}$ is defined as

\begin{equation}
\textrm{SI-SDR}(\bm s, \hat{\bm s}) = 10\log_{10}\left(\frac{\|\alpha \bm s\|^2}{\|\alpha \bm s - \hat{\bm s}\|^2}\right).
\end{equation}

\noindent where $\alpha = \textrm{argmin}_{\alpha}\|\alpha s - \hat{s}\|^2 = \hat{s}^\top s/\|s\|^2$. Specifically, for each mixture on the test set we use the ground truth direction of each active source $\bm \theta_k$ as the target direction for all methods. Then, the separation performance is computed between the ground truth source signal $\gt_k$ and the estimated signal $\estimated(\doa_k)$. \added{The SI-SDR is a metric used to evaluate the quality of separated audio sources by measuring the signal-to-noise ratio between the ground truth signal and the estimated signal, \added{which may have} an arbitrary scaling factor. An SI-SDR of 0 dB signifies that the power of the distortion is equal to the power of the ground truth signal. A positive SI-SDR value indicates that the ground truth signal has more power than the distortion, while a negative SI-SDR signifies that the power of the distortion is greater than that of the ground truth signal. Hence, higher SI-SDR values are desired.}

Furthermore, we assess the performance of the models to predict silent regions in the directions where no source\added{s} \deleted{is}\added{are} placed, i.e., its spatial selectivity. To this end, we introduce the \added{sources-to-silence ratio} (SSR). This measure is also independent of the scaling of the predicted signals. SSR is defined as

\added{
\begin{equation}
\textrm{SSR}_k(\hat{\bm s}) = 10\log_{10}\left( \frac{ \frac{1}{K} \sum_k \| \hat{\bm s}(\doa_k)\|^2} {\frac{1}{I}\sum_t \| \hat{\bm s}(\target)\|^2}\right ), 
\end{equation}
}

\noindent  
where $\{ \target \in  \bm \Theta \|  \angle (\target, \doa_k) \added{>} 2.5^\circ \}$ 
are $I$ directions from a set of \added{36} quasi-uniformly arranged directions on the sphere in a t-design \added{($t = 8$)}, $\bm \Theta$ \cite{hardin_mclarens_1996}, excluding those that are within $2.5^\circ$ of the target direction\added{s}. Note that $\hat{\bm s}(\target)$ and $\hat{\bm s}(\doa_k)$ are the signals predicted for the target directions $\target$ and the ground truth source direction\added{s} $\doa_k$ respectively. \deleted{During testing, for each Ambisonics mixture we average the $K$ SSR values and report the overall median SSR of all mixtures.} \added{Note that a SSR of 0~dB means no spatial selectivity, as it is the case for an omnidirectional receiver. A SSR of $\infty$~dB would be achieved if silence was predicted at all directions that are at least $2.5^\circ$ away from the sources.}

\subsection{Evaluation}

We are interested in evaluating the source separation and spatial selectivity performance of all methods depending on the Ambisonics order. To this end, we compare the SI-SDR and the SSR performance considering Ambisonics mixtures with an order between one and four. In addition, we assess the performance of all methods considering the type of acoustic conditions (anechoic or room) and the type of signals in the Ambisonics mixture (music or universal). \added{All variants are used like a traditional signal-independent beamformer, in that a target direction is provided from which sound shall be extracted, while sound from other directions shall be suppressed. Here, we provide the ground truth direction of the source to all the approaches.} In the following, we report the results and behavior of the evaluated methods considering all conditions. \added{Apart from discussing the numerical values, we also provide visualization of one example, see Figure~\ref{fig:visualization}. Therein, the first maps show the root mean square (RMS) output of the methods, when specifying different target directions. The SI-SDR maps show the SI-SDR for the three sources present in the mix, again, when pointing to different target directions.}

\subsubsection{SH beamformer baseline}

As expected, SH domain beamformers show improved performance in separation and spatial selectivity as the Ambisonics encoding-order increases, with particularly high values for high orders under anechoic conditions. For example in the case of anechoic music signals, the max-$\bm r_\mathrm{E}$ beamformer achieves the best separation performance for third and fourth order with a SI-SDR of 20.27~dB and 25.40~dB respectively (see Table \ref{t:results_musdb}). However, under room conditions, the performance of the beamformer is lower. The difference in performance of SH beamformers between anechoic and room conditions is clearly observed by the max-SDR beamformer performance in both scenarios. Very high values in the anechoic cases are due to the fact that the max-SDR approach typically places zeros in the direction of the other sources, thus cancelling them completely. Note that the max-SDR beamformer performance corresponds to the \deleted{maximally}\added{maximum} possible separation by spatial processing, i.e SH domain beamforming, alone. As discussed in detail below, the network approaches show the largest benefits over SH beamforming under the room condition. This is to be expected because deep learning based approaches can learn to remove sound reflections through non-linear operations. \added{In contrast, the SH beamformer is strictly limited by its spatial processing resolution, only forming weighted linear combinations of the SH signals, see eq. \eqref{eq:bf}. For low orders, linear spatial processing alone does not provide high selectivity which can be seen in the metrics, and also in Figure~\ref{fig:visualization}, where the output of a first order beamformer is shown in the upper right. The output power does not change strongly depending on the target direction.}

\begin{table*}[ht!]
	\centering
	\caption{SI-SDR and SSR median scores in dB\added{, along with their 95\% confidence interval within braces,} calculated for the test set using music signals for both anechoic and room conditions. \added{The highest performances are highlighted using bold font.}}
	\label{t:results_musdb}
	\subcaption*{Musdb18 Anechoic}
	\resizebox{0.90\width}{!}{
	\begin{tabular}{c | c c  c c  c c  c c  c c | c}
		\makecell{\textbf{Ambi.} \\ \textbf{ order}} & \multicolumn{2}{c}{\makecell{\textbf{Demucs} \\ \textbf{(Refinement)}}} & \multicolumn{2}{c}{\makecell{\textbf{Demucs} \\ \textbf{(Implicit)}}} & \multicolumn{2}{c}{\makecell{\textbf{Demucs} \\ \textbf{(Mixed)}}} & \multicolumn{2}{c}{\makecell{\textbf{max-DI} \\ \textbf{BF}}} & \multicolumn{2}{c|}{\makecell{\textbf{max-$\bm r_{\mathrm{E}}$} \\ \textbf{BF}} } & \makecell{\textbf{max-SDR} \\ \textbf{BF} \deleted{(Oracle)}} \\
		 & SI-SDR & SSR & SI-SDR & SSR & SI-SDR & SSR & SI-SDR & SSR & SI-SDR & SSR & SI-SDR  \\ \cline{1-12}
		\multirow{2}{*}{1} & 7.50 & -0.30 & 15.66 & \textbf{10.70} & \textbf{16.45} & 8.79 & 4.48 & 2.71 & 4.40 & 2.48 & 53.83 \\
        & (0.47) & (0.08) & (0.25) & \textbf{(0.21)} & \textbf{(0.28)} & (0.18) & (0.22) & (0.08) & (0.22) & (0.08) & (0.20)\\\cline{1-12}
		\multirow{2}{*}{2} & 14.89 & 0.01 & 15.44 & \textbf{10.91} & \textbf{20.62} & 8.13 & 10.69 & 5.09 & 12.20 & 4.69 & 57.73 \\
        & (0.51) & (0.07) & (0.24) & \textbf{(0.17)} & \textbf{(0.40)} & (0.15) & (0.22) & (0.06) & (0.23) & (0.08) & (0.19)\\\cline{1-12}
		\multirow{2}{*}{3} & 20.16 & 0.16 & 11.85 & \textbf{14.94} & 18.95 & 0.94 & 14.76 & 7.29 & \textbf{20.27} & 6.52 & 59.33 \\
        & (0.54) & (0.08) & (0.12) & \textbf{(0.27)} & (0.47) & (0.04) & (0.22) & (0.04) & \textbf{(0.22)} & (0.06) & (0.20)\\\cline{1-12}
		\multirow{2}{*}{4} & 24.39 & -0.28 & 16.16 & \textbf{10.08} & 23.00 & -0.12 & 17.80 & 9.17 & \textbf{25.40} & 8.31 & 60.29 \\
        & (0.54) & (0.04) & (0.26) & \textbf{(0.13)} & (0.49) & (0.04) & (0.22) & (0.03) & \textbf{(0.23)} & (0.04) & (0.21)\\\cline{1-12}
	\end{tabular}}
	\bigskip
	\subcaption*{Musdb18 Room}
	\resizebox{0.90\width}{!}{%
	\begin{tabular}{c | c c  c c  c c  c c  c c | c}
		\makecell{\textbf{Ambi.} \\ \textbf{ order}} & \multicolumn{2}{c}{\makecell{\textbf{Demucs} \\ \textbf{(Refinement)}}} & \multicolumn{2}{c}{\makecell{\textbf{Demucs} \\ \textbf{(Implicit)}}} & \multicolumn{2}{c}{\makecell{\textbf{Demucs} \\ \textbf{(Mixed)}}} & \multicolumn{2}{c}{\makecell{\textbf{max-DI} \\ \textbf{BF}}} & \multicolumn{2}{c|}{\makecell{\textbf{max-$\bm r_{\mathrm{E}}$} \\ \textbf{BF}} } & \makecell{\textbf{max-SDR} \\ \textbf{BF} \deleted{(Oracle)}} \\
		  & SI-SDR & SSR & SI-SDR & SSR & SI-SDR & SSR & SI-SDR & SSR & SI-SDR & SSR & SI-SDR \\ \cline{1-12}
		\multirow{2}{*}{1} & -1.16 & 0.20 & 0.57 & \textbf{4.90} & \textbf{0.76} & 3.85 & -1.36 & 2.03 & -1.33 & 1.76 & 0.58 \\
        & (0.33) & (0.08) & (0.25) & \textbf{(0.15)} & \textbf{(0.26)} & (0.16) & (0.22) & (0.07) & (0.22) & (0.07) & (0.20) \\\cline{1-12}
		\multirow{2}{*}{2} & 2.94 & 1.14 & \textbf{3.57} & \textbf{6.14} & 2.97 & 0.86 & 1.83 & 3.73 & 1.50 & 3.38 & 4.95 \\
        & (0.38) & (0.09) & \textbf{(0.19)} & \textbf{(0.13)} & (0.34) & (0.10) & (0.22) & (0.07) & (0.23) & (0.08) & (0.19)\\\cline{1-12}  
		\multirow{2}{*}{3} & 4.79 & 1.81 & \textbf{5.31} & \textbf{6.88} & 5.03 & 1.98 & 4.04 & 5.23 & 3.51 & 4.77 & 8.39 \\
        & (0.36) & (0.10) & \textbf{(0.19)} & \textbf{(0.11)} & (0.25) & (0.12) & (0.22) & (0.06) & (0.23) & (0.07) & (0.20)\\\cline{1-12}  
		\multirow{2}{*}{4} & 3.87 & 1.89 & 6.32 & \textbf{7.43} & \textbf{6.43} & 1.43 & 5.53 & 6.45 & 5.07 & 5.93 & 11.05 \\
        & (0.37) & (0.09) & (0.18) & \textbf{(0.13)} & \textbf{(0.21)} & (0.11) & (0.22) & (0.07) & (0.23) & (0.07) & (0.21)\\\cline{1-12}  
	\end{tabular}}
	
\end{table*}

\subsubsection{Refinement mode}

In refinement mode, the deep neural network is used solely for the refinement of the single channel max-$\bm r_\mathrm{E}$ beamformer output, pointing to the target direction. As one would expect, the refinement mode separation performance is closely related to the one achieved by the max-$\bm r_\mathrm{E}$ beamformer. Nevertheless, the refinement mode often improves the max-$\bm r_\mathrm{E}$ beamformer separation. In the cases where the max-$\bm r_\mathrm{E}$ beamformer separation is already high, the refinement mode achieves the best separation performance compared to other operating modes. This is the case for anechoic conditions and universal signals (see Table \ref{t:results_fuss}), where the refinement mode achieves a SI-SDR of 15.76~dB for third order and 21.46~dB for fourth order. However, it underperforms compared \added{to} the other methods under room conditions because the initial separation of the max-$\bm r_\mathrm{E}$ beamformer is not as good. Regarding spatial selectivity, the refinement mode fails to predict silence in source-free regions. The neural network cannot determine if the input sound coming from the max-$\bm r_\mathrm{E}$ beamformer output should be silenced or enhanced without the target direction information. The bad spatial selectivity performance is especially notable in anechoic conditions where, for example, it achieves a SSR of \deleted{-0.23~dB}\added{-0.28~dB} for fourth order when using music signals.

\begin{table*}[ht!]
	\centering
	\caption{SI-SDR and SSR median scores in dB\added{, along with their 95\% confidence interval within braces,} calculated for the free universal sound separation test set (2-4 sources) for both anechoic and room conditions. \added{The highest performances are highlighted using bold font.}}
	\label{t:results_fuss}
	\subcaption*{FUSS Anechoic}
	\resizebox{0.90\width}{!}{
	\begin{tabular}{c | c c  c c  c c  c c  c c | c}
		\makecell{\textbf{Ambi.} \\ \textbf{ order}} & \multicolumn{2}{c}{\makecell{\textbf{Demucs} \\ \textbf{(Refinement)}}} & \multicolumn{2}{c}{\makecell{\textbf{Demucs} \\ \textbf{(Implicit)}}} & \multicolumn{2}{c}{\makecell{\textbf{Demucs} \\ \textbf{(Mixed)}}} & \multicolumn{2}{c}{\makecell{\textbf{max-DI} \\ \textbf{BF}}} & \multicolumn{2}{c|}{\makecell{\textbf{max-$\bm r_{\mathrm{E}}$} \\ \textbf{BF}} } & \makecell{\textbf{max-SDR} \\ \textbf{BF} \deleted{(Oracle)}} \\
		  & SI-SDR & SSR & SI-SDR & SSR & SI-SDR & SSR & SI-SDR & SSR & SI-SDR & SSR & SI-SDR \\ \cline{1-12}
		\multirow{2}{*}{1} & 2.73 & -0.01 & 9.93 & \textbf{8.26} & \textbf{10.20} & 6.61 & 2.47 & 3.39 & 2.61 & 2.99 & 40.94 \\
        &  (0.70) & (0.14) & (0.48) & \textbf{(0.26)} & \textbf{(0.54)} & (0.27) & (0.66) & (0.19) & (0.67) & (0.20) & (0.43) \\\cline{1-12}
		\multirow{2}{*}{2} & 10.99 & -0.01 & 12.27 & \textbf{7.89} & \textbf{15.88} & 7.80 & 7.49 & 6.53 & 9.58 & 5.95 & 42.44 \\
        & (0.83) & (0.15) & (0.47) & \textbf{(0.27)} & \textbf{(0.58)} & (0.28) & (0.64) & (0.22) & (0.73) & (0.23) & (0.44) \\\cline{1-12}
		\multirow{2}{*}{3} & \textbf{15.76} & -0.08 & 13.15 & 7.86 & 15.37 & -0.06 & 11.27 & \textbf{8.83} & 15.28 & 8.19 & 43.39 \\
        & \textbf{(0.82)} & (0.14) & (0.45) & (0.26) & (0.77) & (0.14) & (0.67) & \textbf{(0.21)} & (0.81) & (0.22) & (0.45) \\\cline{1-12}
		\multirow{2}{*}{4} & \textbf{21.46} & -0.10 & 12.98 & 8.94 & 20.22 & -0.13 & 14.91 & \textbf{10.70} & 20.25 & 10.01 & 44.17 \\
        & \textbf{(0.76)} & (0.16) & (0.43) & (0.25) & (0.70) & (0.15) & (0.66) & \textbf{(0.23)} & (0.78) & (0.22) & (0.45)\\\cline{1-12}
	\end{tabular}}
	\bigskip
	\bigskip
	\subcaption*{FUSS Room}
	\resizebox{0.90\width}{!}{%
	\begin{tabular}{c | c c  c c  c c  c c  c c | c}
		\makecell{\textbf{Ambi.} \\ \textbf{ order}} & \multicolumn{2}{c}{\makecell{\textbf{Demucs} \\ \textbf{(Refinement)}}} & \multicolumn{2}{c}{\makecell{\textbf{Demucs} \\ \textbf{(Implicit)}}} & \multicolumn{2}{c}{\makecell{\textbf{Demucs} \\ \textbf{(Mixed)}}} & \multicolumn{2}{c}{\makecell{\textbf{max-DI} \\ \textbf{BF}}} & \multicolumn{2}{c|}{\makecell{\textbf{max-$\bm r_{\mathrm{E}}$} \\ \textbf{BF}} } & \makecell{\textbf{max-SDR} \\ \textbf{BF} \deleted{(Oracle)}} \\
		  & SI-SDR & SSR & SI-SDR & SSR & SI-SDR & SSR & SI-SDR & SSR & SI-SDR & SSR & SI-SDR  \\ \cline{1-12}
		\multirow{2}{*}{1} & -4.22 & 2.01 & -1.50 & \textbf{5.08} & \textbf{-1.27} & 4.65 & -4.08 & 2.36 & -4.17 & 2.08 & -1.75 \\
        & (0.60) & (0.19) & (0.44) & \textbf{(0.23)} & \textbf{(0.48)} & (0.23) & (0.46) & (0.17) & (0.46) & (0.16) & (0.41)\\\cline{1-12}
		\multirow{2}{*}{2} & -0.98 & 2.80 & 1.62 & \textbf{7.16} &  \textbf{2.22} & 3.22 & -0.91 & 4.25 & -1.22 & 3.95 & 2.88 \\
        & (0.63) & (0.22) & (0.36) & \textbf{(0.20)} & \textbf{(0.39)} & (0.21) & (0.44) & (0.19) & (0.44) & (0.19) & (0.35) \\\cline{1-12}
		\multirow{2}{*}{3} & 1.54 & 3.72 & 3.00 & \textbf{7.47} & \textbf{3.75} &  4.01 & 1.59 & 5.76 & 0.99 & 5.37 & 6.40 \\
        & (0.53) & (0.23) & (0.33) & \textbf{(0.22)} & \textbf{(0.37)} & (0.19) & (0.43) & (0.20) & (0.42) & (0.19) & (0.34) \\\cline{1-12}
		\multirow{2}{*}{4} & 3.61 & 4.43 & 3.59 & \textbf{7.62} & \textbf{5.47} &  4.91 & 2.84 & 6.77 & 2.65 & 6.46 & 9.04 \\
        & (0.50) & (0.24) & (0.37) & \textbf{(0.23)} & \textbf{(0.34)} & (0.21) & (0.41) & (0.20) & (0.40) & (0.20) & (0.34)\\\cline{1-12}
	\end{tabular}}
	\bigskip
\end{table*}

\subsubsection{Implicit mode}

\begin{figure*}[ht!]
    \center
    \label{}
    \includegraphics[]{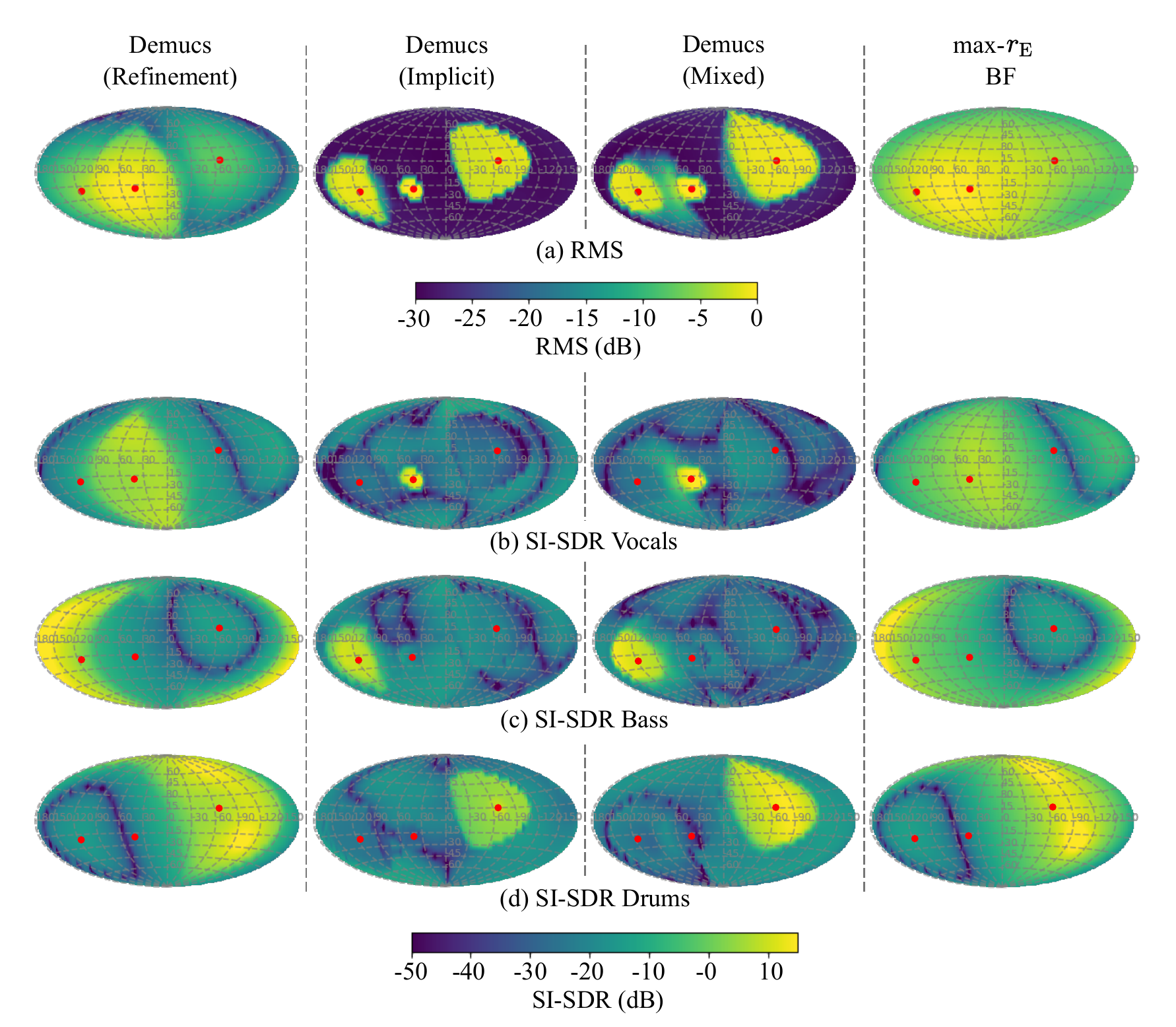}
    \caption{Visualization of a) RMS and b), c), d) SI-SDR for audio predicted in a discrete set of equiangularly distributed directions ($100^\circ \times 50^\circ$) by several methods. The predictions are made using a first order Ambisonics mixture from the Musdb18 test set in anechoic conditions. The red dots correspond to the location of the sources. Corresponding listening examples can be found online \cite{listening_examples}.}
    \label{fig:visualization}
\end{figure*}

In the anechoic case, the implicit mode achieves similar separation and spatial selectivity independent of the encoding order it has been trained and tested on. For music signals, it achieves a SI-SDR of 15.66~dB and a SSR of \deleted{20.66~dB}\added{10.70~dB} for first order mixtures while it achieves a SI-SDR of 16.16~dB and a SSR of \deleted{19.32~dB}\added{10.08~dB} for fourth order mixtures. Hence, low orders are enough for the implicit mode to learn the correspondence between the spatial information contained in the Ambisonics signal and the conditioning direction, and higher order input channels do not have large benefits. The same behaviour is observed when the implicit mode is trained using universal signals. When the implicit mode is trained and tested under room conditions, results show improved performance in separation and spatial selectivity as the encoding order is increased. For instance with music signals, it achieves a SI-SDR of 0.57~dB and a SSR of \deleted{6.50~dB}\added{4.90~dB} for first order mixtures as opposed to 6.32~dB and \deleted{12.70~dB}\added{7.43~dB} achieved for fourth order mixtures (see Table \ref{t:results_musdb}). Overall, the implicit mode is the operating mode that achieves the best SSR results out of all other operating modes for every Ambisonics order and type of signals. \added{This behaviour can also be seen in Figure~\ref{fig:visualization}. The network manages to strongly suppress the signal in directions that are not close to sources.} In addition, it achieves competitive separation performance, especially for music signals under room conditions (see Table \ref{t:results_musdb}).

\subsubsection{Mixed mode}

In the mixed mode, both the first order Ambisonics signal and the beamformer output are provided to the network, while the target direction is also used to condition the network output.
Overall, the mixed mode achieves competitive SI-SDR results compared to all other operating modes. For the more realistic case, i.e. with universal signals and room conditions, the mixed mode achieves the best separation performance compared to all other methods for any Ambisonics order (see Table \ref{t:results_fuss}). For first order musical mixtures in the room, the mixed mode achieves a SI-SDR of 0.76~dB while the max-SDR beamformer achieves a SI-SDR of 0.58~dB. This means that the mixed mode separation performance \added{is similar} to the \deleted{maximally}\added{maximum} possible achieved by spatial processing alone.
Regarding spatial selectivity under room conditions, the mixed mode offers a similar performance independently of the encoding order. For example using universal signals, it achieves a SSR of \deleted{6.09~dB}\added{4.65~dB} for first order and \deleted{6.84~dB}\added{4.91~dB} for fourth order. For anechoic conditions, the SSR values show a\added{n} unexpected behaviour of the model, which performs competitively for lower orders while it fails for higher orders. This can be seen using universal signals in Table \ref{t:results_fuss}, where it achieves a SSR of \deleted{7.89~dB}\added{6.61~dB} for the first order and a SSR of \deleted{-0.45~dB}\added{-0.13~dB} for the fourth order. We suspect that this network behaviour is caused by not learning the correspondence between the target angle and the mixture, and just basing the predictions on the provided max-$\bm r_\mathrm{E}$ beamformer output. This leads to good separation only in anechoic and high-order scenarios, where the max-$\bm r_\mathrm{E}$ beamformer output already provides a meaningful solution according to the training loss function.

\subsubsection{Number of sources in the mixture}

We are also interested in studying the source separation performance depending on the number of sources in the mixture. Based on \deleted{previous}\added{the} results \added{in Table \ref{t:results_fuss}}, we report the performance considering the best learning\added{-}based and SH beamformer methods for universal signals under room conditions, i.e. the mixed mode and the max-DI beamformer.

\begin{figure}[h!]
    \center
    \includegraphics[]{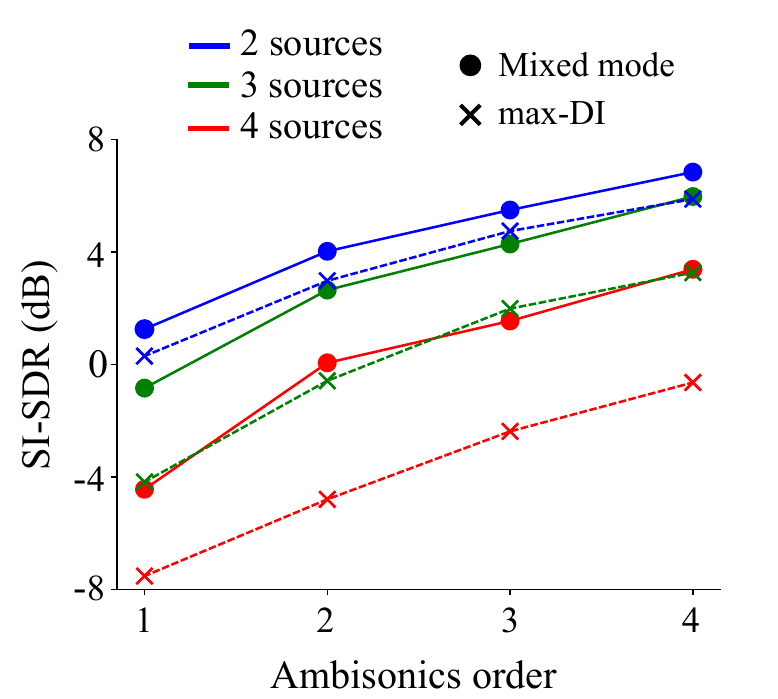}
    \caption{FUSS room SI-SDR median results reported for the different number of sources (2, 3, 4) in the Ambisonics mixture for both mixed mode and max-DI beamformer}
    \label{fig:fuss_room_results_each_number_of_sources}
\end{figure}

Figure \ref{fig:fuss_room_results_each_number_of_sources} shows that for a \deleted{fixed}\added{given} encoding order, the difference between the SI-SDR achieved by the mixed mode compared to the max-DI beamformer increases with the number of sources in the mixture. For a \deleted{fixed}\added{given} number of sources, the improvement of SI-SDR, between the mixed mode and the max-DI beamformer, is more or less constant independently of the encoding order. 
\added{This analysis indicates that as the number of sources within the mixture increases, the utilization of deep learning for source separation becomes more beneficial in comparison to the beamformer approach. However, it should be noted that the increment in performance remains consistent across the Ambisonics encoding orders evaluated.}

\section{Discussion}
\label{sec:discussion}

The results in Section \ref{sec:evaluation} indicate that each of the presented operating modes has its own benefits, depending on the application scenario.

The refinement mode is adequate when the separation achieved by the SH beamformer, which is provided as input to the network, is already high.
For low orders, the refinement mode performs poorly because the SH beamformer does not strongly cancel interfering sources. In this case, the network can not determine which source to refine, as it does not have access to any information concerning which part of the signal is the target, besides level differences. The low SSR values achieved by the refinement mode show that the network is not able to predict silence. It will refine any source at the output of the SH beamformer, even when it is pointing into a source-free region. 

The results achieved by the implicit mode show that the neural network can implicitly learn the correspondence between the spatial information contained in the Ambisonics mixture and the target direction. It correctly interprets the target direction and extracts the sound from that direction. This also becomes evident in the SI-SDR maps shown in \deleted{Fig.}\added{Figure}~ \ref{fig:visualization}, where, for the same input mixture, the neural network predictions are different for each of the conditioning target directions and correlate better with the ground truth signals in the vicinity of the source location.

Under room conditions, the implicit mode also performs better separation than common SH beamformers. This is also true for low orders in anechoic conditions. Interestingly, the implicit mode does not benefit from higher orders in anechoic mixtures. The largest advantage of the implicit mode over conventional SH beamforming lies in the higher spatial selectivity.\deleted{, especially under room conditions.} Pointing to a target direction where no source is present yields very low outputs. \added{This can be seen in the RMS map shown in Figure~\ref{fig:visualization}.} Note that this capability could potentially be used for source localization as well, in an algorithm more similar to \cite{jenrungrot2020cone}, where separation and localization are combined. \added{It also means that the separation stage is more sensitive to direction of arrival mismatches than the SH approaches, i.e., if a sound source is too far away from the provided target direction, the network would output silence. Note that the angular range in which the network outputs signal can be controlled in the training stage, by changing the perturbation applied to the target direction and the minimal distance between sources.} 

Overall, this shows that a source separation network conditioned only on the target direction can perform the whole beamforming operation, i.e. extract sound from specific directions, with mixtures containing an unknown number of sources and arbitrary sound type.

If the goal is to perform source separation where spatial selectivity is less important, the mixed mode might be the most useful. It provides best separation performance for first and second order in the anechoic music dataset and for all orders in the the FUSS dataset under room conditions. Testing the mixed mode with the FUSS dataset under room conditions shows the largest improvements over SH beamforming for relatively complex scenes with four sources. The increase in SI-SDR over the implicit mode comes at the cost of lower SSR, which can also be seen in \deleted{Fig.}\added{Figure}~\ref{fig:visualization}, where the RMS at directions different from the sources is higher. 

Interestingly, neither the implicit mode, nor the mixed mode, which has the max-$\bm r_\mathrm{E}$ beamformer as input,  achieves higher SI-SDR in the anechoic cases than the conventional max-$\bm r_\mathrm{E}$. We see this mainly as evidence for the fact that SH beamforming under anechoic conditions is very effective, and any impairment caused by the network will be enough to result in a lower SI-SDR. The high separation performance of high order conventional SH beamforming is quickly lost in the more realistic case, when room reflections are present. For the first order room condition, the mixed mode achieves higher separation than SH beamformers. \deleted{It even surpasses the max-SDR beamformer,}\added{It is in a similar range as the max-SDR beamformer,} which represents the maximal SDR achievable by frequency-independent SH beamforming given knowledge of the ground truth signal.

In the future, the evaluation of deep learning based methods may be extended to recordings with real microphone arrays. There, the maximal Ambisonics order is not obtained at low frequencies, due to necessary regularization \cite{zotter_ambisonics_2019}. This might have more impact on the performance of conventional approaches compared to deep learning ones, given that the performance of conventional SH beamforming strongly depends on the order. However, it should be tested if a model trained with the data from one microphone array can generalize to data from a \deleted{difference}\added{different} array.

Furthermore, when comparing the objective metrics for the three modes against the SH beamformer, it is important to note that higher separation may come at the cost of time-frequency artifacts, that can occur in the neural network output, but not in SH beamformers. \added{Ideally, a formal listening experiment should be conducted in the future. For now, listening examples are provided online that can give an impression of the perceptual quality \cite{listening_examples}. Although the results are promising, the above artifacts could become audible in a post-processing application. Nevertheless, increasing the level of a particular signal might be feasible, without significantly reducing its quality.}
\deleted{For additional listening examples  demonstrating this trade-off, the reader can refer to the accompanying website.}

\section{Conclusions}
\label{sec:conclusion}
In this paper we proposed the use of end-to-end deep learning as an alternative to conventional SH beamforming. We have shown and analyzed three different operation modes: 1) refinement, 2) implicit and 3) mixed. Specifically, the implicit and mixed mode show that a source separation network can learn associations between a target direction and the information contained in an Ambisonics scene. This allows for using such a network as one would use a beamformer, specifying a target direction and separating arbitrary sounds without adapting the training to a specific number or type of source.

The results show that, under anechoic conditions, the largest separation improvement of the proposed approaches with respect to SH beamforming is achieved for lower Ambisonics orders. In addition, better spatial selectivity is provided for all orders. Under room conditions, the application of deep learning increases both separation and spatial selectivity for all orders. Generally, the behaviour of each operating mode is similar when trained and tested using musical signals or arbitrary signals from a dataset for universal source separation.

\section*{Data availability statement}

Listening examples are available online at: \url{http://research.spa.aalto.fi/publications/papers/acta22-sss/}. 
The code accompanying the paper is provided at: \url{https://github.com/francesclluis/direction-ambisonics-source-separation}.

\section*{Acknowledgements}
This project has received funding from the European Union's Horizon 2020 research and innovation programme under the Marie Sk\l{}odowska-Curie grant agreement No 812719.

\bibliographystyle{IEEEtran}
\bibliography{IEEEabrv,refs}

\end{document}